\documentclass[preprint,12pt]{aastex}

\begin{document}

\shorttitle{GRB Energy Relationships}

\title{Testing the Gamma-Ray Burst Energy Relationships}
\author{David L. Band\altaffilmark{1,2} and
Robert D. Preece\altaffilmark{3}}
\altaffiltext{1}{GLAST SSC, Code 661, NASA/Goddard Space
Flight Center, Greenbelt, MD  20771}
\altaffiltext{2}{Joint Center for Astrophysics, Physics
Department, University of Maryland Baltimore County, 1000
Hilltop Circle, Baltimore, MD 21250}
\altaffiltext{3}{Department of Physics, University of
Alabama in Huntsville, Huntsville, AL  35899}
\email{dband@milkyway.gsfc.nasa.gov,
rob.preece@msfc.nasa.gov}

\begin{abstract}
Building on Nakar \& Piran's analysis of the Amati relation
relating gamma-ray burst peak energies $E_p$ and isotropic
energies $E_{\rm iso}$, we test the consistency of a large
sample of BATSE bursts with the Amati and Ghirlanda (which
relates peak energies and actual gamma-ray energies
$E_\gamma$) relations.  Each of these relations can be
expressed as a ratio of the different energies that is a
function of redshift (for both the Amati and Ghirlanda
relations) and beaming fraction $f_B$ (for the Ghirlanda
relation).  The most rigorous test, which allows bursts to
be at any redshift, corroborates Nakar \& Piran's
result---88\% of the BATSE bursts are inconsistent with the
Amati relation---while only 1.6\% of the bursts are
inconsistent with the Ghirlanda relation if $f_B=1$.  Even
when we allow for a real dispersion in the Amati relation
we find an inconsistency.  Modelling the redshift
distribution results in an energy ratio distribution for
the Amati relation that is shifted by an order of magnitude
relative to the observed distribution; any sub-population
satisfying the Amati relation can comprise at most
$\sim18$\% of our burst sample.  A similar analysis of the
Ghirlanda relation depends sensitively on the beaming
fraction distribution for small values of $f_B$; for
reasonable estimates of this distribution about a third of
the burst sample is inconsistent with the Ghirlanda
relation. Our results indicate that these relations are an
artifact of the selection effects of the burst sample in
which they were found; these selection effects may favor
sub-populations for which these relations are valid.
\end{abstract}


\section{Introduction}

Recently correlations between various energies
characterizing gamma-ray bursts have been reported (Amati
et al. 2002, henceforth A02; Ghirlanda, Ghisellini \&
Lazzati 2004, henceforth G04).  If true, these correlations
have significant implications for burst physics, and could
supplement incomplete observations in compiling burst
databases.  Building on Nakar \& Piran (2005, henceforth
NP05), we test these relations for consistency with a
subset of the bursts observed by the Burst and Transient
Source Experiment (BATSE) that flew on the {\it Compton
Gamma-Ray Observatory (CGRO)}.  Note that consistency would
not prove the validity of these relations.

A02 found that the peak energy $E_p$, the energy of the
maximum of $E^2N(E)\propto \nu f_\nu$ for the entire burst,
and the apparent isotropic energy $E_{\rm iso}$, the total
burst energy if the observed flux were radiated in all
directions, are related: $E_p \propto E_{\rm iso}^{1/2}$.
This is the Amati relation. Subsequently G04 found that
$E_p$ and the actual emitted energy $E_\gamma$ are even
more tightly correlated: $E_p \propto E_\gamma^{0.7}$. This
is the Ghirlanda relation.  In these relations $E_p$ is in
the burst frame, and is found by fitting the spectrum of
the `fluence spectrum,' the spectrum formed from summing
all the burst emission. $E_{\rm iso}$ is calculated from
the observed bolometric energy fluence using the burst
redshift, while $E_\gamma$ is determined from $E_{\rm iso}$
with corrections for the beaming of the gamma-ray emission
derived from analyzing afterglows. These relations were
discovered and calibrated using the small number of bursts
for which afterglows and redshifts have been determined.
The calibrating burst database is small and heterogeneous
enough that the fitted parameters of these relations are
sensitive to the editing of the burst database; for
example, see Friedman \& Bloom (2005, henceforth FB05).  In
addition, the sample is subject to selection effects
resulting from localizing the burst in different wavebands,
tracking the afterglow, and determining the redshift of the
afterglow or the host galaxy.

NP05 pointed out that while larger burst databases lack the
redshifts necessary to calibrate the Amati relation, these
databases can test the validity of this relation because
the ratio $E_p^2/E_{\rm iso}$ cannot exceed a maximum value
for all redshifts.  They found that a large fraction of the
bursts in the databases they considered cannot satisfy the
Amati relation.

We build on the work of NP05 using a database of 760 BATSE
bursts with observed $E_{p,obs}$ and fluences, and test
both the Amati and Ghirlanda relations. To calculate
$E_\gamma$ for the Ghirlanda relation we need the beaming
fraction $f_B$, for which an afterglow must be detected and
monitored.  The most rigorous test assumes $f_B=1$, in
which case the Ghirlanda relation becomes the Amati
relation---$E_{\rm iso} = E_\gamma$ when $f_B=1$---albeit
with a different exponent for $E_p$. We find that indeed a
large fraction of the bursts in our sample are inconsistent
with the Amati relation, as NP05 found, but only a small
fraction are inconsistent with the Ghirlanda relation under
the extreme condition that $f_B=1$.

These rigorous tests of these two relations put bursts at
the redshift that maximizes the $E_p^2/E_{\rm iso}$ and
$E_p^{0.7}/E_\gamma$ ratios, and for the Ghirlanda relation
assumes the extreme value $f_B=1$.  However, we can create
less rigorous yet still relevant tests by comparing the
observed distributions of these ratios to calculated
distributions based on model redshift and beaming fraction
distributions. Applying this model dependent test, we find
that the observed and model distributions are discrepant,
although the discrepancy is less extreme for the Ghirlanda
relation than for the Amati relation.

Redshifts have been calculated for BATSE bursts assuming
empirical relations such as the lag-luminosity (see Band,
Norris \& Bonnell 2004) and $E_p$-luminosity relations
(Yonetoku et al. 2004, henceforth Y04).  We do not use
these redshifts in our analysis here because the result of
that analysis would be a test of the consistency of two
empirical relations, not the validity of a particular
empirical relation.

Similar tests can be applied to relations such as the
$E_p$--$L_B$ relation (Schaefer 2003; Y04; Ghirlanda et al.
2005, henceforth G05a), where $L_B$ is the peak bolometric
luminosity and $E_p$ is the peak energy for either the peak
of the lightcurve (Schaefer) or the fluence spectrum (Y04;
G05a). Since both Y04 and G05a used this relation to
calculate redshifts for a sample of BATSE bursts, finding a
physically reasonable redshift distribution, their relation
implicitly passes the test.

Similar to NP05, we suspect that the Amati relation, and to
a lesser extent the Ghirlanda relation, result from
selection effects affecting the burst sample used to
discover and calibrate this relation. The calibrating
sample falls on the high fluence, low $E_p$ edge of the
BATSE sample. While our results show that these relations
are not valid for the entire BATSE sample, they do not
preclude their validity for a sub-population, although for
the Amati relation this sub-population is small.

In the next section (\S 2) we present the methodology
behind our tests.  The following section (\S 3) describes
our data. We then discuss the results (\S 4) and present
our conclusions (\S 5).

\section{Methodology}

The Amati relation is
\begin{equation}
E_p = C_1
   \left( {{E_{\rm iso}} \over {10^{52}\hbox{ erg}}} \right)^{\eta_1}
\end{equation}
where $E_p$ is the peak energy for the `fluence spectrum'
(the spectrum averaged over the entire burst) in the burst
frame and $E_{\rm iso}$ is the isotropic energy, the burst
energy if the observed flux were emitted in all directions
(A02) . FB05 find $C_1=95\pm 11$~keV and $\eta_1=0.5\pm
0.04$; we use $C_1=95$~keV and $\eta_1=0.5$ for our
calculations.

The Ghirlanda relation is
\begin{equation}
E_p = C_2
   \left( {{E_{\gamma}} \over {10^{51}\hbox{ erg}}} \right)^{\eta_2}
\end{equation}
where $E_\gamma$ is the total energy actually radiated
(G04). FB05 find $C_2=512\pm 15$~keV and $\eta_2 = 0.70\pm
0.07$; we use $C_2=512$~keV and $\eta_2=0.7$.

The peak energy in the observer's frame is
\begin{equation}
E_{p,obs} = E_p / (1+z)
\end{equation}
where $z$ is the burst's redshift.  The total energy
radiated is
\begin{equation}
E_\gamma = E_{\rm iso} \left(1-\cos\theta_j\right) =
   f_B E_{\rm iso}
\end{equation}
where $\theta_j$ is the jet opening angle and $f_B$ is the
beaming fraction, which is determined observationally from
modelling the evolution of the afterglow. The isotropic
energy is
\begin{equation}
E_{\rm iso} = {{4\pi S_\gamma d_L^2}\over{1+z}}
\end{equation}
where $S_\gamma$ is the bolometric fluence and $d_L$ is the
luminosity distance.

Consequently, the Amati relation implies
\begin{equation}
\xi_1 = {{E_{p,obs}^2}\over {S_\gamma}} =
   {{4\pi d_L^2 C_1^2}
   \over{\left[10^{52}\hbox{ erg}\right](1+z)^3}} = A_1(z)
   \quad ,
\end{equation}
and the Ghirlanda relation implies
\begin{equation}
\xi_2 = {{E_{p,obs}^{1.429}}\over {S_\gamma}} =
   f_B {{4\pi d_L^2 C_2^{1.429}}
   \over{\left[10^{51}\hbox{ erg}\right](1+z)^{2.429}}}
   =f_B A_2(z)
   \quad.
\end{equation}
Since $f_B=\left(1-\cos\theta_j\right)$ ranges between 0
and 1, $A_2(z)$ is the upper limit to the $\xi_2$ ratio. We
use $\xi_1$ and $\xi_2$ as the observed ratios and $A_1(z)$
and $A_2(z)$ as theoretical functions of $z$.

Assuming $\Omega_m = 0.3$, $\Omega_\Lambda=0.7$, and $H_0 =
70$~Mpc/km/s, Figure~1 shows $A_1(z)$ (dashed curve) and
$A_2(z)$ (solid curve), respectively. As NP05 pointed out,
observed values of $\xi_1$ that exceed the maximum value of
$A_1(z) = 1.1 \times 10^{9}$ cannot satisfy the Amati
relation (note that NP05 scaled $\xi_1$ by $8\times
10^{-10}$). Similarly, observed values of $\xi_2$ that
exceed the maximum value of $A_2(z) = 2.9 \times 10^{10}$
cannot satisfy the Ghirlanda relation. Since $A_1(z)$ and
$A_2(z)$ are both 0 at $z=0$, both ratios do not have
useful lower bounds.  Note that in reality the two
relations may have a dispersion that may not result from
errors in the measured energies, and the maximum values of
the ratios may be slightly greater, as discussed below.

Note that if $\xi_1$ does not exceed the maximum value of
$A_1(z)$ then eq.~1 can be solved for $z$.  In general, if
an empirical relation between burst characteristics (e.g.,
the Amati, Ghirlanda or Yonetoku relations) results in a
physically reasonable redshift distribution when the
relation is applied to a burst dataset without measured
redshifts (e.g., the dataset we use), then that relation
implicitly passes the type of test used in this paper.

These tests are rigorous but understate any inconsistency
between the data and the proposed relations.  $A_1(z)$ and
$A_2(z)$ peak at $z=3.8$ and $z=12.6$, respectively, both
redshift values greater than the redshifts where most
bursts are currently found to originate.  In addition, the
jet opening angle is generally small such that $f_B =
\left(1-\cos\theta_j\right) \simeq \theta_j^2/2 \ll 1$. The
maximum values of $A_1(z)$ and $A_2(z)$ provide absolute
tests of whether observed values of $\xi_1$ and $\xi_2$ can
possibly satisfy the Amati and Ghirlanda relations,
respectively.  However, we construct model-dependent tests
that compare the observed distributions of $\xi_1$ and
$\xi_2$ with distributions resulting from convolving
$A_1(z)$ and $A_2(z)$ with models of the burst redshift and
beaming fraction distributions.

Thus, if $p(z)\propto dN/dz$ is the redshift probability
distribution then
\begin{equation}
P(>\xi_1)_{\rm model} = \int dz\, p(z)
   H\left[ A_1(z)-\xi_1 \right] \quad ,
\end{equation}
where $H(x)$ is the Heaviside function, 1 for $x\ge 0$ and
0 otherwise.  The burst rate is assumed to follow the star
formation rate, and we use $p(z)$ based on the star
formation rate in Rowan-Robinson (2001); we found the same
qualitative results using the star formation rates
parameterized in Guetta, Piran \& Waxman (2004) based on
the rates of Rowan-Robinson (1999) and Porciani \& Madau
(1999).  Our conclusions are the same for the burst rate
found by Schaefer, Deng, \& Band (2001), which keeps rising
as the redshift increases; Watanabe et al. (2003) finds
that the gamma-ray background can be explained if star
formation follows this rate.

Next, if $p(f_B)$ is the probability distribution for $f_B$
then
\begin{equation}
P(>\xi_2)_{\rm model} = \int df_B \,
   p(f_B)\int dz\, p(z)
   H\left[ f_B A_2(z)-\xi_2 \right] \quad.
\end{equation}
Note that we assume that the beaming fraction does not
evolve with redshift.  As we discuss below, the results are
sensitive to $p(f_B)$.

\section{Data}

We use a sample of 760 BATSE bursts for which we have both
spectral fits to their `fluence spectra' and energy
fluences. The 16 channel CONT spectra of bursts between
April 1991 and August 1996 (the 4th BATSE
catalog---Paciesas et al. 1999) with sufficient fluxes
were fit with a number of different spectral models
(Mallozzi et al. 1998); we use the $E_{p,obs}$ from the
fits with the `Band' function (Band et al. 1993).

The fluences are from the 4th BATSE catalog (Paciesas et
al. 1999).  The catalog presents 20--2000~keV fluences,
which we treat as bolometric.   Bloom, Frail \& Sari (2001)
showed that the k-correction for burst spectra is of order
unity, and FB05 use a k-correction to shift the fluence's
energy band to 20--2000~keV in the burst frame (which
requires the burst redshift).  Jimenez, Band \& Piran
(2001) showed that the fluences resulting from fitting high
resolution spectra and from the processing pipeline used to
create the BATSE catalog can differ by up to a factor of 2.
This provides a measure of the uncertainty in the fluence
and indicates that attempts to extend the energy band of
the fluence using the spectral fits is unnecessary.

\section{Results}

Figure~1 shows that the maximum value of the ratio
$E_{p,obs}^2/ S_\gamma=1.1\times10^9$ occurs at $z=3.825$.
For our BATSE burst database, 668 out of 760 bursts, or
88\% of the bursts, exceed this maximum.  Even if we
increase this maximum value by a factor of 2 (5) to account
for a real dispersion around the Amati relation and
uncertainties in the determination of $E_{p,obs}$ and
$S_\gamma$, 555 (391) out of 760 bursts, or 73\% (51\%) of
the bursts, exceed this increased maximum. By this rigorous
test most of our bursts are inconsistent with the Amati
relation. Ghirlanda, Ghisellini \& Firmani (2005,
henceforth G05b) quantify the deviation from the Amati
relation of the bursts in an edited version of our database
in terms of the dispersion around the Amati relation of the
calibrating bursts (i.e., bursts with spectroscopic
redshifts).  G05b claim that there is no inconsistency with
the Amati relation, yet Nakar \& Piran (2005, private
communication) point out that the probability of finding
the number of bursts whose ratio $\xi_1$ exceeds the
limiting value by a given factor is negligible.

On the other hand, Figure~1 also shows that the maximum
value of the function $A_2(z) = \xi_2 / f_B  = 2.85\times
10^{10}$ occurs at $z=12.577$. Assuming $f_B=1$, only 12
out of 760 bursts, or 1.6\% of the bursts, violate the
Ghirlanda relation.  If the ratio's maximum value is
doubled (or even increased five-fold) to account for a real
dispersion around the relation and uncertainties in the
observations, then only 6 out of 760 bursts, or 0.8\% of
the bursts violate this increased maximum.

Comparisons of the model and observed distributions of the
energy ratio $\xi_1$ for the Amati relation (Figure~2)
shows that the observed distribution is shifted to larger
values by an order of magnitude.  This reinforces our
conclusion above that our data and the Amati relation are
inconsistent.

The analysis of the Ghirlanda relation requires not only
the redshift distribution but also the beaming fraction
distribution; as shown by Figure~3, we find that the model
distribution for the ratio $\xi_2$ is heavily dependent on
the assumed beaming fraction distribution.  Very small
beaming factors result in afterglows with very early breaks
in the afterglow evolution, before observations have begun,
while large beaming factors lead to afterglows that break
late, after the afterglow is no longer observable.
Consequently the currently observed distribution of beaming
factors is plagued by selection effects at both its low and
high ends; these selection effects are not relevant to the
BATSE database but they are largely the same effects that
shaped the datasets in which the two relations were
discovered. Nonetheless, Figure~3 compares the observed
distribution (solid curve) to the model distribution
resulting from three different estimates of the beaming
fraction distribution (dashed curves). The first uses the
{\it observed} $f_B$ distribution found by Frail et al.
(2001): $p(f_B)\propto f_B^{-1.77}$ for $\log(f_B)>-2.91$
and constant for smaller $f_B$.  The second beaming
fraction distribution is based on Guetta et al. (2004), who
modelled the actual $\theta_j$ distribution as a steep
power law above $\theta_{j,0}=0.12$ radians, and a much
shallower power law for smaller $\theta_j$. Transforming
$\theta_j$ into $f_B$ using the small angle approximation
$f_B \simeq \theta_j^2/2$, and converting the actual
distribution into the observed distribution (a burst with a
beaming factor of $f_B$ has a probability of $f_B$ of being
observed) results in $p(f_B)\propto f_B^{-2}$ for
$\log(f_B)>-2.14$ and constant for smaller $f_B$.  Note
that Guetta et al. perform a more sophisticated conversion
(relying on additional modelling assumptions) from the
actual to observed distributions taking into account the
burst luminosity function and the distance to which bursts
can be detected.  The differences between these two beaming
faction distributions led us to fit the values of
$\theta_j$ in FB05, resulting in the third distribution on
Figure~3: $p(f_B)\propto f_B^{-1.65}$ for $\log(f_B)>-2.41$
and $p(f_B)\propto f_B^{0.7}$ for smaller $f_B$.  For
comparison, the dot-dashed curves result from constant
values of $f_B=0.0275$ and $f_B=1$. Although $f_B=0.0275$
maximizes the Kolmogorov-Smirnoff (K-S) probability that
the model and observed distributions are drawn from the
same population, this probability is still only $1.11\times
10^{-8}$.

The first three model distributions shown on Figure~3
result from beaming factor distributions with similar power
law indices for large beaming fractions but break values
that differ significantly. Power law distributions with
indices $\mu<-1$ (where $p(f_B)\propto f_B^\mu$) diverge as
$f_B$ approaches 0, and the value of the normalized
probability at a given value of $f_B$ above any break or
cutoff depends on the value of this break or cutoff. The
magnitude of the discrepancy between the observed and model
distributions of $\xi_2$ differ for the three estimates of
the beaming fraction distribution, but in all cases there
is a real discrepancy.

Perhaps the BATSE bursts consist of a number of different
burst populations, only one of which satisfies the Amati or
Ghirlanda relations. To test this hypothesis for each
relation, we progressively removed the highest ratio from
the observed distribution and calculated the K-S
probability that the resulting observed and model
distributions are the same. Thus for the Amati relation we
first calculated the K-S probability that the observed
distribution of the ratio $\xi_1 = E_{p,obs}^2/ S_\gamma$
(the solid curve in Figure~2) was drawn from a model
distribution (the dashed curves in Figure~2). We then
sorted the observed ratio distribution. Iteratively, we
removed the highest ratio, and calculated the K-S
probability.  We performed the same procedure for the
Ghirlanda ratio $\xi_2$ for the three estimates of the
beaming fraction distribution.

Table~1 shows the K-S probabilities for the total burst
population and the sub-population that maximizes the K-S
probability; the table also shows the fraction that this
sub-population constitutes.  The small K-S probabilities
for the full sample for each relation quantifies the
discrepancies between the observed and model distributions,
which are readily apparent from Figures~2 and~3.  No more
than $\sim18$\% of the burst sample is from a
sub-population satisfying the Amati relation.  Whether a
sub-population can satisfy the Ghirlanda relation depends
on the beaming factor distribution; a larger break in the
beaming factor distribution increases both the size of the
sub-population that may be consistent with the Ghirlanda
relation and the K-S probability of this sub-population.

The redshift distribution used above assumes that BATSE
detected bursts with equal efficiency at all redshifts.
However, identical bursts originating at higher redshift
will be fainter and their observed spectra will be softer
than their low-$z$ counterparts.  Assuming there is no
compensating evolution, the detection efficiency for higher
redshift bursts should be smaller, and the observed
redshift distribution should be shifted to smaller
redshifts. Indeed, this appears to be the case for a sample
of bursts observed by BATSE that were binned with respect
to intensity (Mallozzi et al. 1995). There is a strong
general trend toward smaller average $E_p$ for the weaker
bursts, and the interpretation is consistent with a larger
average cosmological redshift.  The model redshift
distribution assuming redshift-independent detection
efficiency peaks at $z=1.7$, which is below the redshifts
where the Amati and Ghirlanda ratios peak (the ratios shown
in Figure~1).  Thus shifting the observed redshift
distribution to lower $z$ can only shift the model ratio
distributions (the dashed curves in Figures~2 and~3) to
lower values, increasing the discrepancy between the model
and observed distributions.

\section{Conclusions and Summary}

We test whether the Amati relation (A02)---$E_p \propto
E_{\rm iso}^{1/2}$---and the Ghirlanda relation
(G04)---$E_p\propto E_{\gamma}^{0.7}$---are consistent with
a large sample of BATSE bursts for which $E_{p,obs}$ and
fluences are available.  Note that $E_p$ is in the burst
frame and $E_{p,obs}$ in the Earth's frame. In the most
rigorous test, where the bursts may be at any redshift and
have any beaming fraction for which these relations are
satisfied, we find that the Amati relation cannot be
satisfied by 88\% of the bursts, consistent with the
results of NP05, while the Ghirlanda relation is not
satisfied by only 1.6\% of the bursts.  Accounting for a
real dispersion on these relations does not alter this
conclusion.

A less rigorous test results from modelling the redshift
distribution for both relations and the beaming fraction
$f_B$ distributions for the Ghirlanda relation.  The model
distributions of the ratios $E_{p,obs}^2/ S_\gamma$ for the
Amati relation and $E_{p,obs}^{1.43} / S_\gamma$ for the
Ghirlanda relation are shifted to smaller values of these
ratios relative to the observed distribution.  The
magnitude of the discrepancy for the Ghirlanda relation
depends on the model for the beaming factor distribution,
specifically on breaks or cutoffs in the assumed power law
model at small values of $f_B$.  We use three different
model distributions based on the small number of $f_B$
values, although we note that the set of bursts with $f_B$
is the same set used to discover and calibrate the two
relations, and thus are affected by the same selection
effects, which are not relevant to the BATSE bursts.

Including a detection efficiency that decreases as the
redshift increases exacerbates this discrepancy (unless
there is compensating burst evolution). If we assume that
the Amati or Ghirlanda relations apply to a sub-population
of the entire dataset, then only $\sim18$\% of the BATSE
burst sample can be members of this sub-population for the
Amati relation, whereas the sub-population's size depends
on the beaming factor distributions for the Ghirlanda
relation.

The methodology developed here can also be applied to the
$L_B\propto E_p^{\eta_3}$ relation (Schaefer 2003; Y04;
G05a), where $L_B$ is the bolometric peak luminosity.
Schaefer finds that $\eta_3 = 2.778\pm 0.2$ when he uses
$E_p$ at the time of the peak flux; he does not normalize
the relation.  Y04 find $\eta_3 = 1.94 \pm 0.19$ and
normalize the relation; G05a find similar values. Both Y04
and G05a use $E_p$ for the fluence spectrum; it is peculiar
that quantities for the maximum flux and the entire burst
should be so closely correlated. The relevant ratio is
$P\langle E \rangle/ E_{p,obs}^{\eta_3}$, where $P$ is the
peak photon flux and $\langle E \rangle$ is a
characteristic energy (see Band et al. 2004). This ratio
has a peak value at $z\simeq 5$ for Schaefer's value of
$\eta_3=2.778$, and rises monotonically for $\eta_3=1.94$
found by Y04.  There is no need to test this ratio for the
Y04 relation since both Y04 and G05a use this relation to
calculate reasonable redshifts for BATSE bursts.

These results suggest that these two relations may be
artifacts of selection effects in the burst sample in which
these relations were discovered.  The selection effects may
favor a burst sub-population for which the Amati or
Ghirlanda relation is valid.  Bursts for which redshifts
and beaming fractions have been determined must be
relatively bright and soft (low $E_{p,obs}$, the energy
range in which the localizing instruments operate) to be
localized and for their afterglows to be tracked. Figure~4
shows the distribution of our BATSE burst sample in the
$E_{p,obs}$--fluence plane and the bursts from the FB05
sample; also seen are the limits resulting from the Amati
(solid curve) and Ghirlanda (assuming $f_B=1$; dashed
curve) relations.  We converted the fluences of the FB05
sample to the 20 to 2000~keV band using the spectral fits
in their paper; where they do not report spectral indices
we used low and high energy spectral indices of $\alpha =
-0.8$ and $\beta=-2.3$, respectively (Preece et al. 2000).
Note that although the BATSE and FB05 fluences are chosen
to be bolometric, in reality they integrate the spectrum
between different energy limits, and they result from
different types of processing.  As can be seen, the bursts
used to calibrate these two relations (i.e., the FB05
sample) are on the edge of the BATSE distribution,
consistent with the Amati relation. As NP05 concluded, the
sample of bursts with redshifts and afterglow observations
have a much higher selection threshold than the BATSE
distribution.

The much larger sample of bursts with redshifts, fluences
and beaming fractions that Swift will collect will test
these relations conclusively, may distinguish between
different burst populations and may reveal truly universal
relations among burst properties.

\acknowledgements

We thank Ehud Nakar and Tsvi Piran for their careful
reading of our text and their insightful comments.  We also
thank Don Lamb for an insightful discussion of the Amati
relation.

\clearpage

\begin{figure}
\plotone{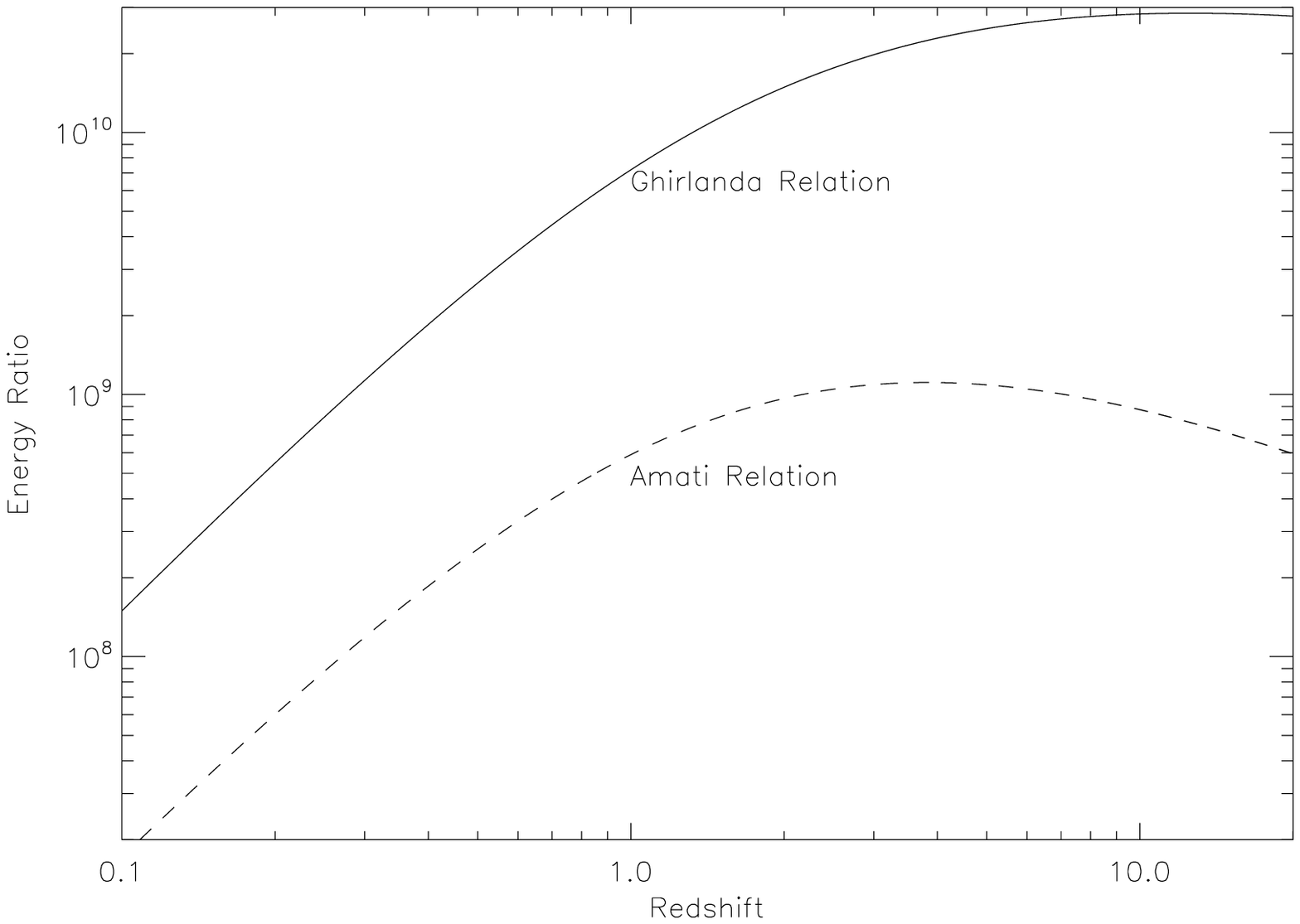} \caption{Predicted value of the Amati
relation energy ratio $E_{p,obs}^2/ S_\gamma$ (dashed
curve) and Ghirlanda relation energy ratio
$E_{p,obs}^{1.43} / [f_B S_\gamma]$ (solid curve) as a
function of redshift.}
\end{figure}

\begin{figure}
\plotone{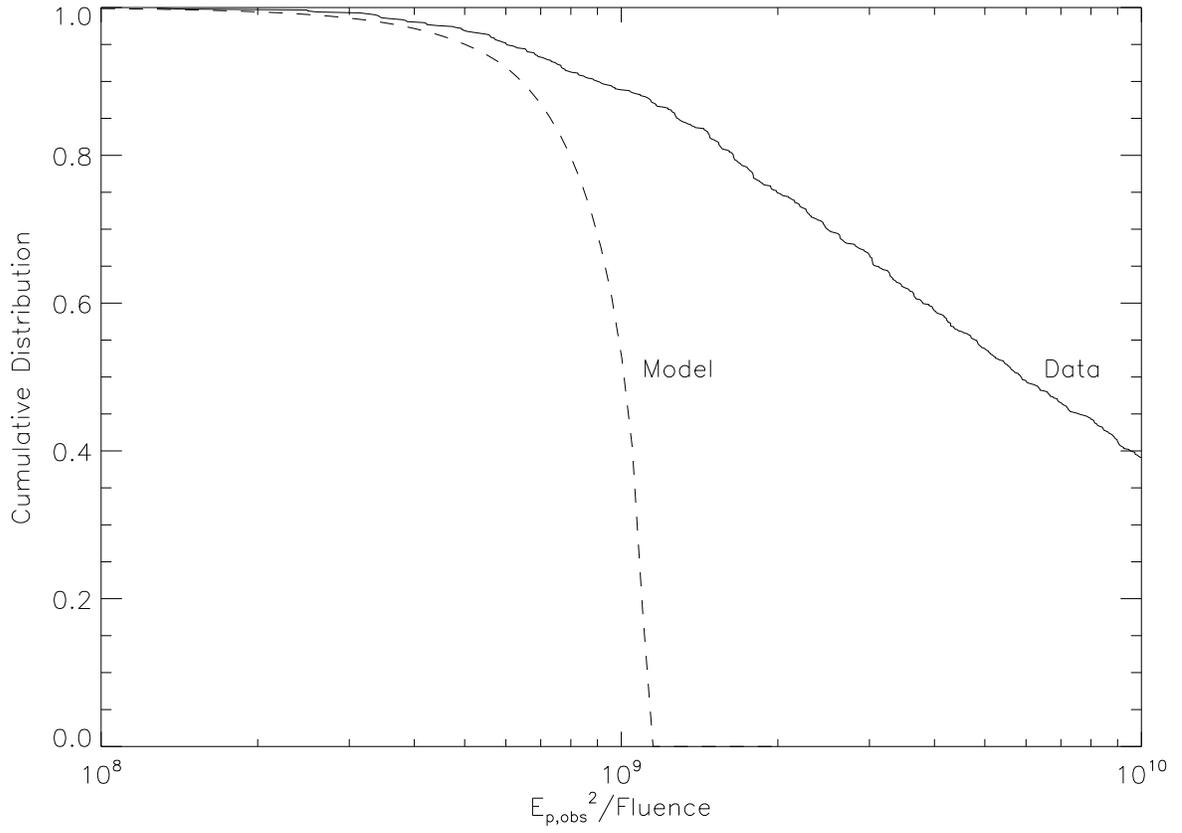} \caption{Observed (solid curve) and model
(dashed curve) distributions of the ratio $E_{p,obs}^2/
S_\gamma$.  If the Amati relation and the assumed redshift
distribution are both valid then these ratio distributions
should be the same.}
\end{figure}

\begin{figure}
\plotone{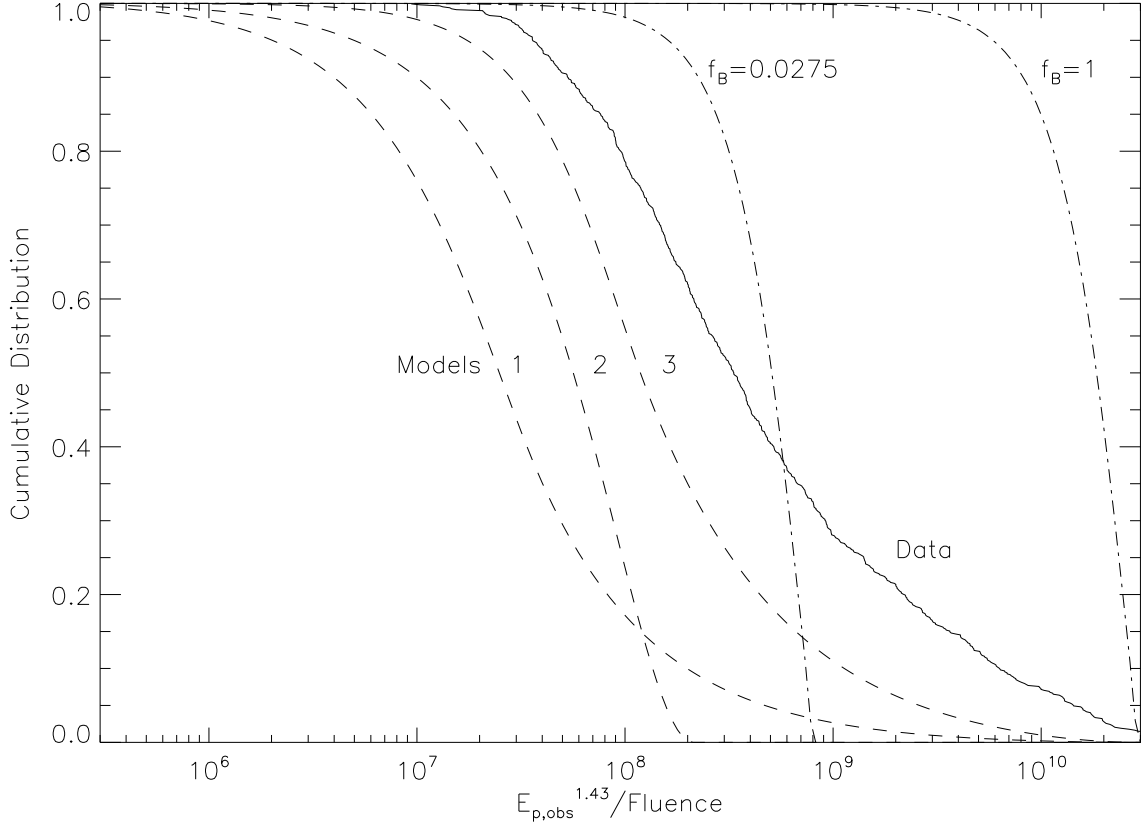} \caption{Observed (solid curve) and model
(dashed curves) distributions of the Ghirlanda relation's
energy ratio $E_{p,obs}^{1.43} / [f_B S_\gamma]$.  The
dashed curves result from the beaming fraction
distributions of Frail et al. (2001; labelled 1), Guetta et
al. (2004; labelled 2) and our fit (labelled 3). If the
Ghirlanda relation and the assumed redshift and beaming
fraction distributions are all valid then the model and
observed ratio distributions should be the same.  For
comparison, the two dot-dashed curves show the ratio
distribution if $f_B=0.0275$ (left) or $f_B=1$ (right).}
\end{figure}

\begin{figure}
\plotone{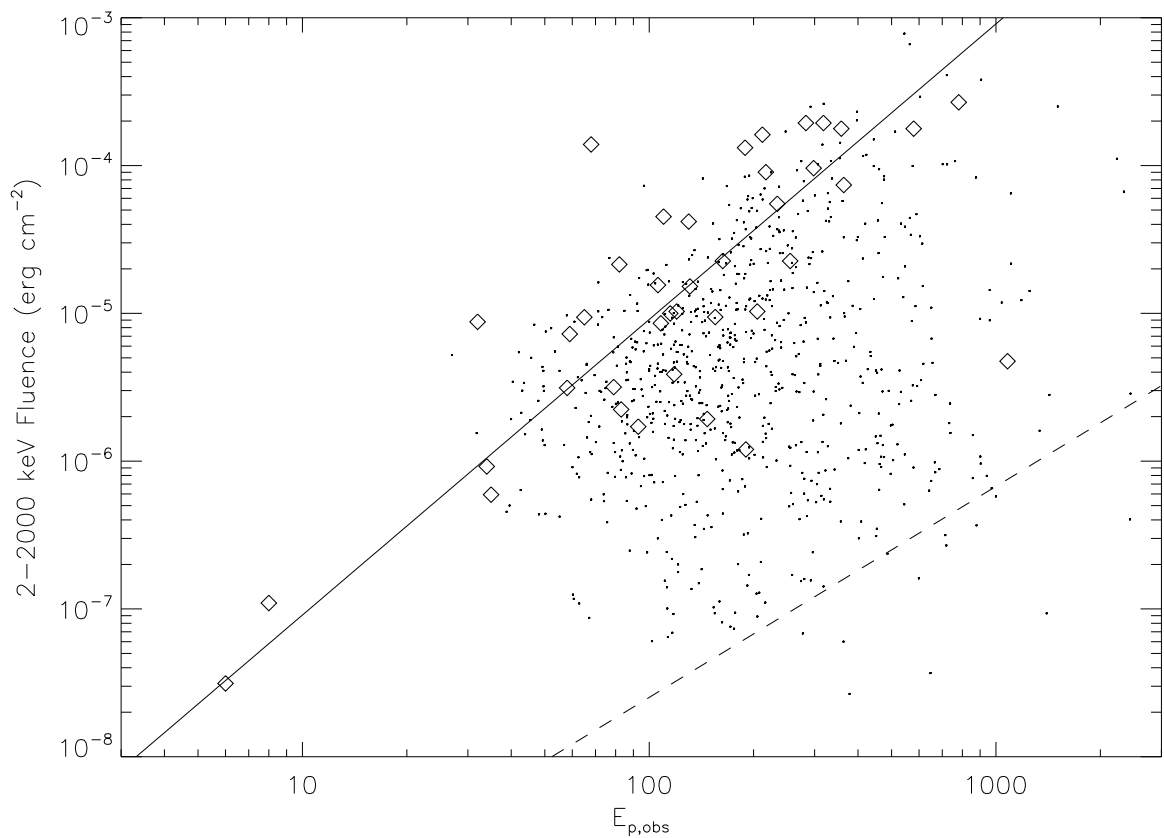} \caption{Location of bursts in the BATSE
(dots) and Friedman \& Bloom (2004---diamonds) samples on
the $E_{p,obs}$--fluence plane.  Also shown are the limits
of the Amati (solid line) and Ghirlanda (for $f_B=1$;
dashed line) relations; these relations permit bursts to
fall above these lines.  Note that the bursts Friedman \&
Bloom use to calibrate the Amati relation have a dispersion
around this relation, yet the BATSE bursts are clearly
fainter and harder (higher $E_{p,obs}$) than the Friedman
\& Bloom bursts.}
\end{figure}

\clearpage

\begin{deluxetable}{l l c c c}
\tablecolumns{5}
\tablecaption{\label{Table1}Consistency Between Observed
and Model Ratio Distributions}
\tablehead{ \colhead{Relation} & \colhead{Beaming Fraction}
& \colhead{K-S Probability,} & \colhead{K-S Probability,} &
\colhead{Subpopulation} \\
\colhead{ } & \colhead{Distribution\tablenotemark{a}} &
\colhead{Entire Population} &
\colhead{Subpopulation\tablenotemark{b}} &
\colhead{Fraction\tablenotemark{c}} }
\startdata
Amati & NA\tablenotemark{d} & $<10^{-38}$ & $1.1\times10^{-4}$ & 0.182 \\
Ghirlanda & Frail et al. (2001) & $5.09\times10^{-33}$
   & $>0.01$\tablenotemark{e} & 0.018\tablenotemark{e} \\
Ghirlanda & Guetta et al. (2004) & $2.69\times 10^{-8}$ & 0.236 & 0.614 \\
Ghirlanda & This paper & $1.56\times 10^{-6}$ & 0.0761 & 0.658 \\
\enddata
\tablenotetext{a}{Estimate of the beaming fraction
distribution.}
\tablenotetext{b}{K-S probability for the
sub-population that maximizes this probability.}
\tablenotetext{c}{Fraction of the 760 BATSE bursts in the
sub-population that maximizes the K-S probability.}
\tablenotetext{d}{The beaming fraction is not required for
the Amati relation.}
\tablenotetext{e}{The K-S probability is above 1\% for 14 bursts in our
sample.}

\end{deluxetable}

\end{document}